\documentclass{elsart}

\usepackage{amssymb}

\begin{document}

\begin{frontmatter}



\title{Superbursts and long bursts as surface phenomenon of
compact objects.}


\author[1]{Monika Sinha},
\author[1]{Mira Dey},
\author[2]{Subharthi Ray},
\author[3]{Jishnu Dey\corauthref{cor1}}
\corauth[cor1]{present address: Department of Physics, Presidency
College, 86/1 College Street, Kolkata 700073, India}

\address[1]{Dept. of Physics, Presidency
College, 86/1 College Street, Calcutta 700 073, India}
\address[2]{Inter University Centre
for Astronomy and Astrophysics, Post bag 4, Ganeshkhind, Pune
411007, India }
\address[3]{Azad Physics Centre, Dept. of Physics, Maulana Azad College,
Calcutta 700 013, India}

\begin{abstract}
We suggest that superbursts from some low mass X-ray binaries may
be due to breaking and re-formation of diquark pairs, on the
surface of realistic strange stars. Diquarks are expected to
break up due to the explosion and shock of the thermonuclear
process. The subsequent production of copious diquark pairing may
produce sufficient energy to produce the superbursts.
\end{abstract}

\begin{keyword}
dense matter \sep elementary particles: diquarks \sep stars:
superburst

\PACS 14.65.-q \sep 95.85.Nv \sep 96.60.Rd \sep 97.80.Jp \sep
98.70.Qy
\end{keyword}
\end{frontmatter}

\section{Introduction}
Type-I X-ray bursts from Low Mass X-ray Binary systems are
believed to be due to thermonuclear fusion. The duration of such
bursts is typically of the order of seconds to minutes. Recently
some such Type-I X-ray bursters show bursts 1000 times longer in
duration and 1000 times more energetic than typical type-I X-ray
bursts. That is why they are known as superbursts. Till date
superbursts have been observed from eight different sources. Two
of them showed repeated superbursts. It should be mentioned that
recently repeated superbursts have been observed from the source
GX 17+2 with luminosity near Eddington luminosity \cite{intz}
while all other sources have luminosity $\sim~(0.1-0.3)\times
L_{Edd}$, $L_{Edd}$ being the Eddington luminosity.

There are models of superbursts for neutron stars in terms of
unstable carbon burning \cite{sb,cb}. But there is disparity
between the superburst energies and recurrence time observed and
the energies and recurrence time predicted by theoretical models.

We have employed the Realistic Strange Star (ReSS) model
\cite{d98sp} to get a good estimate of the large amount of energy
liberated in a superbursts. We suggest that diquarks present on
the ReSS surface are expected to break up due to the explosion and
shock of the thermonuclear processes. The subsequent production
of copious diquark pairing may liberate sufficient energy to
produce the very long bursts observed.

\section{The spin-spin potential and interaction energy}

The quark-quark interaction has a spin dependent component which
can be obtained either from one-gluon exchange between quarks or
from the instanton induced interaction. This part of the
potential is of delta function range which can be transformed to
a smeared potential by introducing the idea of either a finite
glue-ball mass or a secondary charge cloud screening as in
electron-physics \cite{bhaduri}.

The form of the potential is given below :
\begin{equation}
{\rm
H}_{ij}~=~-\frac{2\alpha_s\sigma^3}{3m_im_j\pi^{1/2}}(\lambda_i.\lambda_j)({\rm
S}_i.{\rm S}_j)e^{-\sigma^2 r_{ij}^2}. \label{diq}
\end{equation}
The factor $\sigma^3/\pi^{1/2}$ normalizes the potential. In this
equation $\alpha_s$ is the strong coupling constant, and the
subscripted $m, \lambda$ and $S$ are the constituent masses,
colour matrices and spin matrices for the respective quarks.

For this spin dependent interaction quarks on the ReSS surface
will form diquarks.

   For $N-\Delta$ mass difference (i.e. in $u$-$d$ sector)
Dey et al. \cite{dd} found that this gives $\sigma$ varying from
6 to 2.03 $fm^{-1}$ for a set of $\alpha_s$ 0.5 to 1.12. The
parameters we have used  are given in Table~\ref{tabparam}.

\begin{table}
\caption{Parameters of the Gaussian Potential} \vskip 1cm
\begin{center}
\begin{tabular}{|c|c|c|c|c|c|c|}
\hline Sets & 1&2&3&4&5&6 \\
\hline $\alpha_s$& 0.5 & 0.5&0.87& 0.87& 1.12&1.12 \\
\hline $\sigma (fm^{-1})$& 6.0& 4.56& 0.87& 2.61& 6.0& 2.03 \\
\hline
\end{tabular}
\end{center}
\label{tabparam}
\end{table}

Anti-symmetry of flavour  symmetric di-quark wave function
requires that while space part is symmetric, di-quark must be
either in spin singlet and colour symmetric ($6$) state, or in
spin triplet and colour anti-symmetric ($\bar 3$) state. In both
cases spin-spin force is repulsive\footnote{Private
communication, R. K. Bhaduri.} and formation of pair is inhibited.

For flavour anti-symmetric di-quarks, however, the situation is
the opposite. Colour symmetric $(6)$ configuration is associated
with the spin triplet so that $(\lambda_i.\lambda_j)(S_i.S_j) =
1/3 $ and colour anti-symmetric state ($\bar 3$) goes with the
spin singlet which gives $(\lambda_i.\lambda_j)(S_i.S_j) = 2$.
With overall negative sign in the potential (Eq.\ref{diq}) these
channels produce attraction. Hence there is a probability for
example of $u$, $d$ quarks to pair up predominantly in spin
singlet state. The effect of this can be found easily in our model
since we know the distribution of the $u$ and $d$ quarks in the
momentum space and their Fermi momenta are uniquely determined
from precise and lengthy calculations satisfying beta stability
and charge neutrality.

In addition to spin-colour contribution the potential
(Eq.\ref{diq}) is evaluated in the momentum space. Thus on average
the contribution of a pair can be found as:
\begin{equation}
-\frac12(\lambda_i.\lambda_j)({\rm S}_i.{\rm S}_j)~
\frac{2\alpha_s\sigma^2\pi}{3 m_i
m_j}~\frac{n_u+n_d}{n_un_d}~\frac{6\times2~}{(2\pi)^4}~I.
\label{cor}
\end{equation}
where  $n_u$ and $n_d$ are number density of $u$ and $d$ quarks
respectively at the star surface.

Here
\begin{equation}
I~=~\int_0^{k_{fu}}~\int_0^{k_{fd}}~\int_{-1}^1~f(
k_u,k_d,\theta)~ k_d^2~ dk_d~k_u^2~ dk_u~d(cos\theta_d)
\label{dipot}
\end{equation}

and

\begin{equation}
f(k_i,k_j,\theta)=\frac{1 - {\rm exp}\left(-\frac{\frac{k_i^2 +
k_j^2}{4}-\frac{k_i k_j
cos(\theta_{ij})}{2}}{\sigma^2}\right)}{\frac{k_i^2 +
k_j^2}{4}-\frac{k_i k_j cos(\theta_{ij})}{2}}
\end{equation}

where subscripted $k$'s are momenta of interacting quarks and
$k_{fi}$'s are Fermi momenta for $i$-th species of quark at star
surface.

These correlation energies for different sets of parameters (see
Table \ref{tabparam}) are given in Table \ref{corr}.

\begin{table}[h]
\caption{Integrated values for the pairing energy Eq.(\ref{diq})
for different pairs for spin singlet (colour $\bar 3$) states in
MeV. For spin triplet (colour $6$) state the energies will be six
times less. } \vskip 1cm
\begin{center}
\begin{tabular}{|c|c|c|c|c|c|c|}
\hline  Sets & 1&2&3&4&5&6 \\
\hline pairing energy&-23.578&-23.287&-41.025&-38.225&-52.814&-46.636 \\
\hline
\end{tabular}
\end{center}
\label{corr}
\end{table}

\section{Conclusions}

The interaction producing a coloured diquark in spin zero state,
for example, is a strong one and its overall effect is lowering
of energy by a few $MeV$. Once the pairs are misaligned during
normal burst their recombination may provide bursts over several
hours with energy release estimated to be large. The estimated
total energy liberated, $10^{42}~ ergs$, can be explained in our
model with the calculated pair density $\sim ~0.27/fm^3$ and a
surface thickness of only a micron, if the entire surface is
involved.

It is intriguing to surmise that the elusive properties of some of
the most compressed objects in nature namely the compact stars,
showing superbursts, may be accounted for by the spin alignment of
pairs of the smallest components of matter, $-$ namely the quarks.




\begin{thebibliography}{00}

\bibitem{intz} J. J. M. in 't Zand, R. Cornelisse \& A. Cumming, Astron. \&
Astrophys. {\bf426}, 257 (2004).

\bibitem{sb} T. E. Strohmayer \& E. F. Brown, Astrophys. J.
{\bf566}, 1045 (2002).

\bibitem{cb} A. Cumming \& L. Bildsten, Astrophys. J. Lett. {\bf559}, 127
(2001).

\bibitem{d98sp} M. Dey, I. Bombaci, J. Dey, S. Ray \& B. C.
Samanta, Phys.  Lett. B {\bf 438}, 123 (1998) ; Addendum {\bf
447}, 352 (1999) ; Erratum {\bf 467}, 303 (1999).

\bibitem{bhaduri} R. K. Bhaduri, L. E. Cohler \& Y. Nogami, Phys.
Rev. Lett. {\bf44}, 1369 (1980).

\bibitem{dd} J. Dey \& M. Dey, Phys. Lett.  B {\bf138}, 200 (1984).






\end{thebibliography}
\end{document}